\documentclass[aps,prl,twocolumn,showpacs,preprintnumbers,amsmath,amssymb]{revtex4}
\usepackage{amsmath,amsfonts,amssymb,latexsym}
\usepackage{epsfig}

\begin{document}
\title{Biased random walks on complex networks: the role of local navigation rules}
\author{Agata Fronczak and Piotr Fronczak}
\affiliation{Faculty of Physics and Center of Excellence for
Complex Systems Research, Warsaw University of Technology,
Koszykowa 75, PL-00-662 Warsaw, Poland}
\date{\today}

\begin{abstract}
We study the biased random walk process in random uncorrelated
networks with arbitrary degree distributions. In our model, the
bias is defined by the preferential transition probability, which,
in recent years, has been commonly used to study efficiency of
different routing protocols in communication networks. We derive
exact expressions for the stationary occupation probability, and
for the mean transit time between two nodes. The effect of the
cyclic search on transit times is also explored. Results presented
in this paper give the basis for theoretical treatment of the
transport-related problems on complex networks, including
quantitative estimation of the critical value of the packet
generation rate.
\end{abstract}  \maketitle


The problem of wandering at random in a network (or lattice) finds
applications in virtually all sciences \cite{bookMazo,bookHavlin}.
With only minor adjustments random walks may represent thermal
motion of electrons in a metal, or migration of holes in a
semiconductor. The continuum limit of the random walk model is
known as diffusion. It may describe Brownian motion of a particle
immersed in a fluid, as well as heat propagation, bacterial
motion, and even fluctuations in the stock market. Recently, the
concept of random walks has been also applied to explore traffic
in complex networks. The spectrum of network related problems
include, among many others, ordinary traffic in a city,
distribution of goods and wealth in economies, biochemical and
gene expression pathways, and finally search (or routing)
strategies in the Internet and other communication networks
\cite{PREAdamic01,AdvTadic02,PhysATadic04,PREKim02,
PRERosvall05,PREWang06,PREGemano06}.

In this paper, we deal with biased random walks on complex
networks, and we explore the role of different local navigation
rules on the mean first-passage (or transit) time between any pair
of nodes \cite{bookRedner}. The biased random walk model defined
on scale-free networks is particularly interesting since it has
been considered as a mechanism of transport and search in real
networks, including the Internet. For a long time one has believed
that the most optimum transport-related processes are based on
shortest paths between two nodes under consideration. At the
moment, one has however understood that such a routing strategy
would require a global knowledge on network topology, which is
often not available. Moreover, one can simply imagine that routing
strategies based on shortest paths may create inconvenient queue
congestions in scale-free networks, given that the majority of the
shortest paths pass through hub nodes in such structures. It has
been also realized that a possible alternative is to consider
local navigation rules instead of global knowledge. As a
consequence, a number of adequate models have been proposed (see
e.g. \cite{PhysATadic04,PREWang06}). In general, the models mimic
traffic in complex networks by introducing packets (particles)
generation rate, as well as assigning a randomly selected source
and a random destination to each packet. In these models, a common
observation is that the traffic exhibits continuous phase
transition from free flow to the congested phase as a function of
the packet generation rate. In the free flow state, the numbers of
created and delivered particles are balanced, while in the jammed
state, the number of packets accumulated in the network increases
with time. In this paper we show that the random walk model,
although very simple, correctly describes properties of the
proposed traffic models in the free flow state. We calculate
transit times characterizing this state. We also give some
suggestions how to calculate the critical packet generation rate.

Technically, we define our random walks as follows. We consider
random uncorrelated networks with given node degree distributions
$P(k)$ \cite{PRENewman01}. The networks are also known as random
graphs or configuration model, and they have been repeatedly shown
to be very useful in modelling different phenomena taking place on
networks. We assume that the networks are connected, i.e. there
exists a path between each pair of nodes. Given the graph
structure, the diffusing particle (packet) is created at a
randomly selected node, and it is assigned a random destination
node. In the next time steps the particle passes from a node to
one of its neighbors being directed by local navigation rules. In
practice, it means that being in a certain node $i$ random walker
performs a local search in its neighborhood (up to the first,
second, or further orders) looking if the destination node is
within the search area. If the destination is found, the particle
is delivered directly to the target following the shortest path
(the rule is known as the cyclic search \cite{PhysATadic04}).
Otherwise, the particle continues biased random walk, i.e. the
next position (a node $j$) is chosen according to the prescribed
probability $w_{ij}$.

In the following, to explore transit times characterizing biased
random walks in uncorrelated networks with arbitrary degree
distributions $P(k)$, we partially reproduce and generalize
standard calculations for the mean first passage time in periodic
lattices \cite{bookHughes}. At the beginning, we work out some
general concepts related to biased random walks without the cyclic
search. In particular, we calculate the stationary occupation
probability $P_i^\infty$ for the diffusing particle, which
describes the probability that the particle is located at the node
$i$ in the infinite time limit. Then, performing simple textbook
calculations we derive formulas for the mean transit time between
any pair of nodes (we would like to stress that some time ago
similar calculations were done for unbiased random walks on
complex networks \cite{PRLNoh04}; results presented in our paper
encompass the results of Ref.~\cite{PRLNoh04} as a special case).
The role of the cyclic search on transit times is explored via a
simple renormalization trick applied to nodes' degrees.

Thus, let us consider a particle that hops at discrete times
between neighboring nodes of a random network described by the
adjacency matrix $\mathbf{A}$. Let $P_{ij}(t)$ be the probability
that the particle starting at site $i$ at time $t=0$ is at site
$j$ at time $t$. The evolution of this occupation probability is
given by the master equation
\begin{equation}\label{MasterEq}
P_{ij}(t+1)=\sum_{l=1}^NA_{lj}w_{lj}P_{il}(t),
\end{equation}
where the meaning of $w_{lj}$ was already exposed, whereas
$A_{lj}$ represents element of the adjacency matrix, which is
equal to $1$ if there exists a link between $l$ and $j$, and $0$
otherwise. In the rest of the paper we perform a detailed analysis
of the local navigation rules defined by the preferential
transition probability \cite{PREWang06,PREYang05}
\begin{equation}\label{wlj}
w_{lj}=\frac{k_j^{\alpha}}{\sum_{m=1}^{k_l}k_m^\alpha},
\end{equation}
where the sum in the denominator runs over the neighbors of the
node $l$, and the exponent $\alpha$ is the model free parameter.
Note that according to the formula (\ref{wlj}) the transition
probability from $l$ to $j$ in our biased random walk depends only
on the connectivity of the next-step node $j$. Note also that for
$\alpha=0$ we recover the ordinary unbiased random walk studied by
Noh and Rieger \cite{PRLNoh04}.

\begin{figure*}
\centerline{\epsfig{file=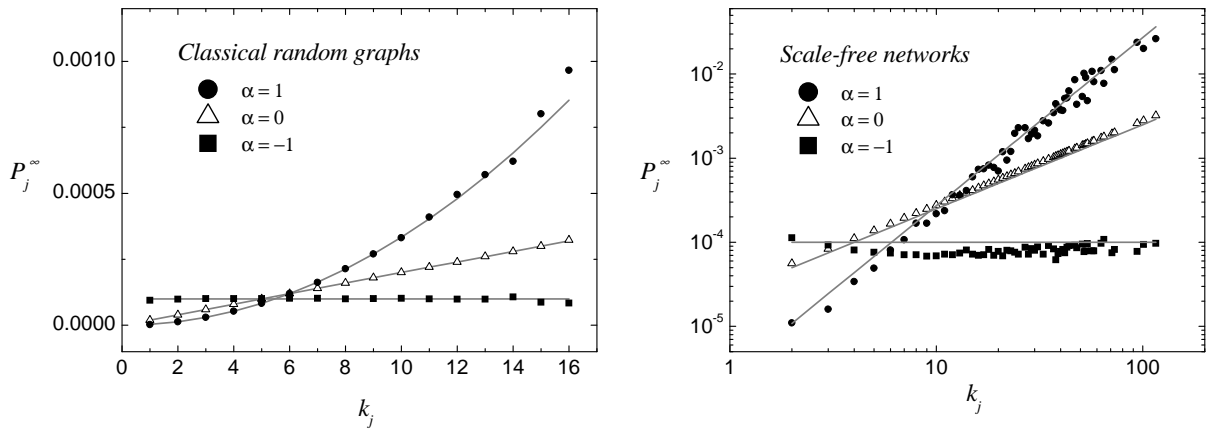,angle=0,width=16cm}}
\caption{Stationary probability distributions $P_j^\infty(k)$
calculated for different values of the parameter $\alpha$ in
classical random graphs and scale-free networks. Solid lines
correspond to the theoretical prediction of Eq.~(\ref{Pinfty}).
All numerical simulations have been done for networks of size
$N=10^4$ and averaged over $10^4$ random walkers. In the case of
classical random graphs $\langle k\rangle=5$ was assumed, whereas
in scale-free networks $\gamma=3$ and $k_{min}=2$ were
chosen.}\label{figPi}
\end{figure*}

In order to calculate the stationary occupation probability
$P_i^\infty$ characterizing the studied biased random walks we
average the master equation (\ref{MasterEq}) over the ensemble of
the considered networks (i.e. we apply mean field approximation to
this equation)
\begin{equation}\label{MasterMF}
P_{j}^\infty\simeq\sum_{l=1}^N\langle A_{lj}\rangle\langle
w_{lj}\rangle P_l^\infty.
\end{equation}
Now, before we proceed further, let us recall a few structural
properties of uncorrelated networks with a given node degree
distribution. First, one can show that probability of a link
between any pair of nodes $l$ and $j$ with degrees respectively
equal to $k_l$ and $k_j$ is given by (see Eq.~(27) in
\cite{PREFronczak06})
\begin{equation}\label{Alj}
\langle A_{lj}\rangle=\frac{k_lk_j}{\langle k\rangle N}.
\end{equation}
Second, since in uncorrelated networks the node degree
distribution $Q(k_m/k_l)$ of the nearest neighbors of a node $l$
does not depend on $k_l$ (compare Eq.~(1) in \cite{AIPFronczak05},
and Eq.~(4) in \cite{handbookNewman})
\begin{equation}\label{Q}
Q(k_m/k_l)=\frac{k_m}{\langle k\rangle}P(k_m),
\end{equation}
the normalization factor in the formula (\ref{wlj}) is equal to
\begin{equation}\label{notePF}
\sum_{m=1}^{k_l}k_m^\alpha=k_l\sum_{m=1}^{k_l}k_m^\alpha
Q(k_m/k_l)=\frac{\langle k^{\alpha+1}\rangle}{\langle
k\rangle}k_l,
\end{equation}
and the transition probability $w_{lj}$ between $l$ and $j$
averaged over different network realizations may be written as
\begin{equation}\label{wljsr}
\langle w_{lj}\rangle=\frac{\langle k\rangle}{\langle
k^{\alpha+1}\rangle k_l} k_j^\alpha.
\end{equation}
Finally, inserting the relations (\ref{Alj}) and (\ref{wljsr})
into the simplified master equation (\ref{MasterMF}), after some
algebra, one obtains
\begin{equation}\label{Pinfty}
P_j^\infty=\frac{k_j^{\alpha+1}}{N\langle k^{\alpha+1}\rangle}.
\end{equation}
Note that for $\alpha=0$, which stands for the unbiased random
walk, the stationary distribution is, up to normalization, equal
to the degree of the the node $j$, i.e. $P_j^\infty\sim k_j$. It
means that the more links a node has, the more often it will be
visited by a random walker. Note also that for $\alpha=-1$, which
represents the anti-preferential transition probability
$w_{lj}\sim 1/k_j$, the stationary occupation probability is
uniform $P_j^{\infty}=1/N$.

To test the validity of Eq.~(\ref{Pinfty}) we have numerically
calculated the fraction of random walkers found in nodes with a
given connectivity $k_j$. The expected power law relation
$P_j^\infty\sim k_j^{\alpha+1}$ was found in all the considered
$\alpha$-cases, and for different classes of the analyzed networks
(i.e. classical random graphs, and scale-free networks $P(k)\sim
k^{-\gamma}$ with the characteristic exponent $\gamma=3$), see
Fig.~\ref{figPi}. The same scaling behavior was found in Ref.
\cite{PREWang06} for the number of packets moving simultaneously
on BA networks \cite{BAnet} in the free flow state. In the
mentioned paper, a packet routing strategy based on the
preferential transition probability (\ref{wlj}), and the so-called
path iteration avoidance, which means that no link can be visited
twice by the same packet, has been considered. At each time step
$R$ packets have been generated in the network, and a fixed node
capacity $C$, that is the number of packets a node can forward to
other nodes, has been assumed. The fact that our results coincide
with those of Wang et al. \cite{PREWang06} shows that packets may
be considered as non-interacting particles (i.e. independent
biased random walkers) in the free flow, stationary state. One can
also show that the approach may be used to estimate the critical
value of packets generation rate $R_c$ \cite{Fronczak07}, as the
parameter should fulfill a kind of balance equation between the
node's processing efficiency $C$, and the number of delivered
packets delivered $P_j^\infty R_c\langle T_{ij}\rangle$, where
$\langle T_{ij}\rangle$ stands for the mean first-passage time
(\ref{Mfpt0}) averaged over all pair of nodes, and respectively
$R_c\langle T_{ij}\rangle$ corresponds to the total number of
packets distributed over the whole network.

\begin{figure*}
\centerline{\epsfig{file=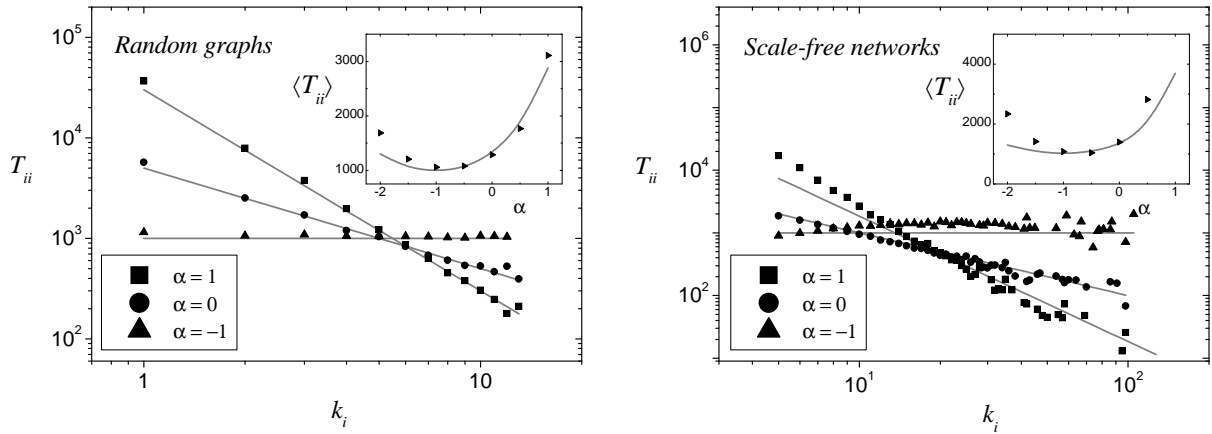,angle=0,width=16cm}}
\caption{Mean first return time $T_{ii}$ vs node degree $k_i$
(main panels), and $\langle T_{ii}\rangle$ vs $\alpha$ (insets) in
classical random graphs and scale-free networks. Numerical
calculations have been done for networks of size $N=10^3$.
$\langle k\rangle=5$ and $\langle k\rangle=2k_{min}=10$ have been
assumed in classical random graphs and scale-free networks (with
$\gamma=3$), respectively. Data presented in the figure have been
averaged over $10^3$ random walkers running in $10^2$ different
network configurations.}\label{figTii}
\end{figure*}

The first-passage probability $F_{ij}(t)$, namely the probability
that the random walk starting at the node $i$ visits $j$ for the
first time at time $t$ satisfies the well-known convolution
relation \cite{PRLNoh04,bookRedner}
\begin{equation}\label{FptEq}
P_{ij}(t)=\delta_{t0}\delta_{ij}+\sum_{\tau=0}^tP_{jj}(t-\tau)F_{ij}(\tau).
\end{equation}
The delta function term in the last equation (\ref{FptEq})
accounts for the initial condition that the walk starts at $i=j$.
Applying the Laplace transform, defined as
$\widetilde{f}(s)=\sum_{t=0}^\infty e^{-st}f(t)$, to this equation
leads to the fundamental expression
\begin{equation}\label{FptLT}
\widetilde{F}_{ij}(s)=\frac{\widetilde{P}_{ij}(s)-\delta_{ij}}{\widetilde{P}_{jj}(s)},
\end{equation}
in which the Laplace transform of the first-passage probability
$\widetilde{F}_{ij}(s)$ is determined by the corresponding
transform of the probability distribution $\widetilde{P}_{ij}(s)$.
Consequently, due to the fact that all moments
\begin{equation}\label{Rijn}
R_{ij}^{(n)}=\sum_{t=0}^\infty
t^n\left(P_{ij}(t)-P_j^\infty\right)
\end{equation}
of the exponentially decaying part of $P_{ij}(t)$ are finite,
expanding $\widetilde{P}_{ij}(s)$ as a power series in $s$
\begin{equation}\label{PijLTseries}
\widetilde{P}_{ij}(s)=\frac{P_j^\infty}{1-e^{-s}}+
\sum_{n=0}^{\infty}(-1)^nR_{ij}^{(n)}\frac{s^n}{n!},
\end{equation}
and then inserting (\ref{PijLTseries}) into (\ref{FptLT}), and
again expanding the result as a series in $s$, one finally obtains
the formula for the mean transit time between $i$ and $j$
\begin{eqnarray}
T_{ij}&
=&\sum_{t=0}^{\infty}tF_{ij}(t)=-\widetilde{F}^{'}_{ij}(0)\nonumber
\\\label{Mfpt0}&=&\left\{
\begin{array}{ll}1/P_j^\infty, & \quad \mbox{ for } j=i \\ [3mm]
\left[ R^{(0)}_{jj}- R^{(0)}_{ij} \right]/P_j^\infty, & \quad
\mbox{ for } j\neq i \end{array} \right. \ .
\end{eqnarray}
At the moment, let us remind that $P_j^\infty$ (\ref{Pinfty})
corresponds to the stationary occupation probability, which has
been already calculated.

\begin{figure*}
\centerline{\epsfig{file=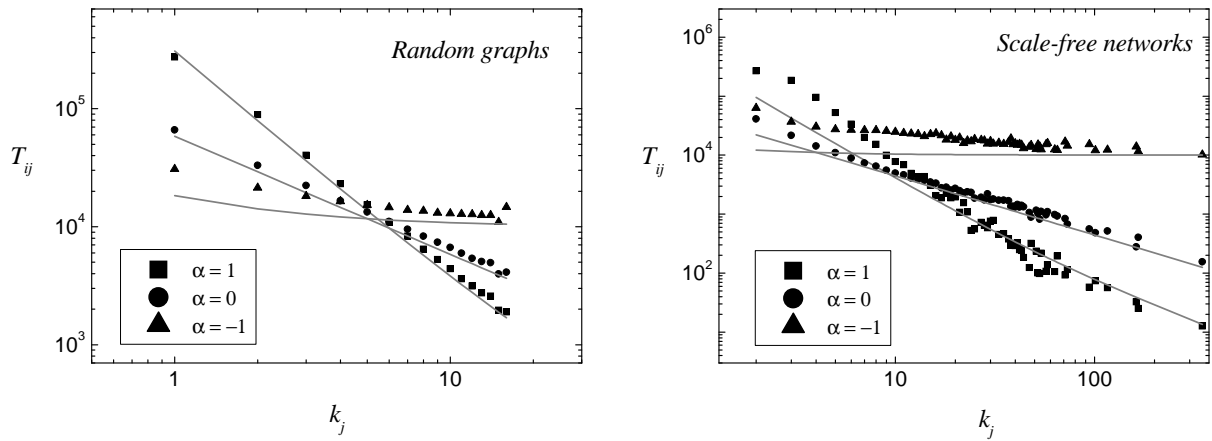,angle=0,width=16cm}}
\caption{Mean transit time $T_{ij}$ between two nodes $i$ and $j$
vs connectivity of the target node $k_j$ in classical random
graphs and scale-free networks. Numerical calculations have been
done for networks of size $N=10^4$. $\langle k\rangle=5$ and
$\langle k\rangle=2k_{min}=4$ have been assumed in classical
random graphs and scale-free networks (with $\gamma=3$),
respectively. Data presented in the figure have been averaged over
$10^2$ random walkers running in $50$ different network
configurations.}\label{figTij}
\end{figure*}

Figure \ref{figTii} shows how the mean first return time $T_{ii}$
of the biased diffusing particle wandering in random network
depends on $k_i$. In the figure, numerically calculated transit
times are indicated by scattered points, whereas their values
predicted by the theory (\ref{Mfpt0}), namely
\begin{equation}\label{Tii}
T_{ii}=\frac{N\langle k^{\alpha+1}\rangle}{k_i^{\alpha+1}},
\end{equation}
are represented by solid lines. Subsets given in the figure show
how the mean first return time $\langle T_{ii}\rangle$ averaged
over all nodes depends on $\alpha$ (i.e. on local navigation rules
governing the diffusing particle)
\begin{equation}
\langle T_{ii}\rangle=N\langle k^{\alpha+1}\rangle\langle
k^{-\alpha-1}\rangle,
\end{equation}
and they indirectly show how fast the biased random walk is. The
minimum value of $\langle T_{ii}\rangle$ observed for
$\alpha_m\simeq-1$ in classical random graphs indicates that the
anti-preferential transition probability (\ref{wlj}) causes the
slowest exploration of the considered networks, which, in turn,
causes that in the case of such a navigation rule the relaxation
part of the occupation probability $P_{ii}(t)-P_i^\infty$
converges to zero much slower then in the case of other values of
the parameter $\alpha$ (the same reasoning applies to the case of
$\alpha_m\simeq-0.5$ in scale-free networks). The reasoning
implies that although, in general, the parameters $R_{jj}^{(0)}$
and $R_{ij}^{(0)}$ in the formula (\ref{Mfpt0}) for the mean
transit time $T_{ij}$ can not be easily calculated, the expected
for $|\alpha-\alpha_m|\neq 0$ fast convergence of the occupation
probability $P_{ij}(t)$ towards the stationary distribution
$P_j^{\infty}$ allows one to simplify the relation
\begin{eqnarray}\nonumber
T_{ij}&\simeq&\left(\sum_{t=0}^2(P_{jj}(t)-P_j^\infty)-\sum_{t=0}^0(P_{ij}(t)-P_j^\infty)\right)/P_j^\infty\\
&=&\frac{N\langle k^{\alpha+1}\rangle}{k_j^{\alpha+1}}+
\frac{N\langle k\rangle^2}{\langle
k^2\rangle}\frac{1}{k_j}-2\label{Tij}.
\end{eqnarray}
In figure \ref{figTij} one can see that the theoretical prediction
of the last equation (\ref{Tij}) quite good agrees with numerical
calculations of $T_{ij}$. As expected, the approximate formula
(\ref{Tij}) works better for the parameter $\alpha>\alpha_m$. We
have also checked that the mean first passage time $T_{ij}$
between any pair of nodes does not depend on the source node $i$
in the considered networks. It only depends on the destination
node $j$.

Knowing the mean transit time (\ref{Mfpt0}) of the biased random
walk, the role of the cyclic search on the quantity can be
calculated through simple renormalization trick applied to nodes'
degrees. The trick consists in dividing the walk between any pair
of nodes $i$ and $j$ into two parts. The first part, before the
diffusing particle hits neighborhood of the target, and the second
part, when the particle sees its destination, and follows the
shortest path. Distinguishing between the two parts allows to
treat the first part as an ordinary biased random walk from a node
$i$ to the node $J$ with the renormalized degree $k_J$ equal to
\begin{equation}\label{kJ}
k_{J}=k_j\left(\frac{\langle k^2\rangle}{\langle
k\rangle}-1\right)^x,
\end{equation}
where $\langle k^2\rangle/\langle k\rangle$ represents the average
connectivity of the nearest neighbors in uncorrelated networks,
and $x$ is the parameter describing the depth of the search area.
Now, since the mean transit time of the second part of the walk is
equal to $x$ the mean first passage time characterizing the whole
walk between $i$ and $j$ (i.e. its both parts) is given by the sum
\begin{equation}\label{Tijstar}
T_{ij}^{(x)}=x+T_{iJ}=x+\left[ R^{(0)}_{JJ}- R^{(0)}_{iJ}
\right]/P_{J}^\infty,
\end{equation}
where the quantities $T_{iJ}$, $P_J^\infty$, $R_{JJ}^{(0)}$, and
$R_{iJ}^{(0)}$ apply to the network, in which the original target
node $j$  together with its nearest neighborhood was replaced by a
single node $J$ of degree $k_J$ (\ref{kJ}). In this case, however,
due to difficulties related to the precise calculation, or even
estimation of the parameters $R_{jj}^{(0)}$ and $R_{ij}^{(0)}$,
the direct verification of the validity of the formula
(\ref{Tijstar}) is impossible. However, we have numerically
checked that the mean first passage time $T^{(x)}_{ij}$
characterizing the cyclic search does not depend is the source
node $i$, and in the first approximation it is proportional to
$k_J^{-\alpha-1}$ (compare Eq.~(\ref{Tij})).

In summary, we have studied the biased random walk process in
random uncorrelated networks with arbitrary degree distributions.
In our model, the bias was defined by the preferential transition
probability (\ref{wlj}) (see also another paper on biased
diffusion in random networks \cite{Sood07}). We have calculated
the expression for the stationary occupation probability, and we
have derived formulas for the mean first passage times between any
pair of nodes. The role of the cyclic search on transit times was
explored via a simple renormalization trick applied to nodes'
degrees. We have also shown that the random walk approach can be
used to explain some properties of traffic dynamics in
communication networks. Other traffic-related problems, that can
be solved using the approach include, among many others,
microscopic explanation of the phase transition from free flow to
the jammed phase, and quantitative estimation of the critical
value of the packet generation rate in scale-free networks
\cite{PREWang06}. We leave the problems to our future work
\cite{Fronczak07}.

The work was funded in part by the State Committee for Scientific
Research in Poland under Grant 1P03B04727 (A.F.), and the European
Commission Project CREEN FP6-2003-NEST-Path-012864 (P.F.).


\begin{thebibliography}{6}
\bibitem{bookMazo} R.M. Mazo, {\it Brownian Motion: Fluctuations, Dynamics, and Applications}, Oxford Univ. Press (2002).
\bibitem{bookHavlin} D. ben-Avraham and S. Havlin, {\it Diffusion and Reactions in Fractals and disordered Systems}, Cambridge Univ. Press (2000).
\bibitem{PREAdamic01} L.A. Adamic et al., Phys. Rev. E {\bf 64}, 046135 (2001).
\bibitem{AdvTadic02} B. Tadi\`{c} and G.J. Rodgers, Adv. Complex Systems {\bf 5}, 445 (2002).
\bibitem{PhysATadic04} B. Tadi\`{c} and S. Thurner, Physica A {\bf 332}, 566 (2004).
\bibitem{PREKim02} B.J. Kim et al., Phys. Rev. E {\bf 65}, 027103 (2002).
\bibitem{PRERosvall05} M. Rosvall, P. Minnhagen, and K. Sneppen, Phys. Rev. E {\bf 71}, 066111 (2005).
\bibitem{PREWang06} W.-X. Wang et al., Phys. Rev. E {\bf 73}, 026111 (2006).
\bibitem{PREGemano06} R. Germano and A.P.S. de Moura, Phys. Rev. E {\bf 74}, 036117 (2006).
\bibitem{bookRedner} S. Redner, {\it A guide to first-passage processes}, Cambridge Univ. Press (2001).
\bibitem{PRENewman01} M.E.J. Newman, S.H. Strogatz, and D.J. Watts, Phys. Rev. E {\bf 64}, 026118 (2001).
\bibitem{bookHughes} B.D. Hughes, {\it Random Walks and Random Environments}, (Vol.~1: Random Walks), Clarendon Press, Oxford (1995).
\bibitem{PRLNoh04} J.D. Noh and H. Rieger, Phys. Rev. Lett. {\bf 92}, 118701 (2004).
\bibitem{PREYang05} S.-J. Yang, Phys. Rev. E {\bf 71}, 016107 (2005).
\bibitem{PREFronczak06} A. Fronczak and P. Fronczak, Phys. Rev. E {\bf 74}, 026121 (2006).
\bibitem{AIPFronczak05} A. Fronczak, P. Fronczak, and J.A. Ho\l yst, AIP Conf. Proc. {\bf 776}, 52 (2005).
\bibitem{handbookNewman} M.E.J. Newman, {\it Random graphs as models of networks}, in {\it Handbook of Graphs and Networks}, S. Bornholdt and H.G. Schuster (eds.), Wiley-VCH, Berlin (2003).
\bibitem{BAnet} A.L.Barab\'asi, R. Albert and H. Jeong, Physica A {\bf 272}, 173 (1999).
\bibitem{Fronczak07} P. Fronczak and A. Fronczak, in preparation.
\bibitem{Sood07} V. Sood and P. Grassberger, eprint cond-mat/0703233.





\end{thebibliography}
\end{document}